\documentclass[prl,preprint,showpacs]{revtex4}
\newcommand{\alp}{{\alpha}}
\newcommand{\om}{{\omega}}
\newcommand{\ep}{{\epsilon}}
\newcommand{\bea}{\begin{eqnarray}}
\newcommand{\beq}{\begin{equation}}
\newcommand{\eea}{\end{eqnarray}}
\newcommand{\eeq}{\end{equation}}

\begin{document}
\title
{Choice of a Metric for the Non-Hermitian Oscillator}
\date{\today}
\author{D.P.~Musumbu, H.B.~Geyer and W.D.~Heiss}

\affiliation{Institute of Theoretical Physics and Department of Physics,
University of Stellenbosch, 7602 Matieland, South Africa }

\begin{abstract}
The harmonic oscillator Hamiltonian, when augmented by a non-Hermitian
$\cal{PT}$-symmetric part, can be transformed into a Hermitian
Hamiltonian. This is achieved by introducing a metric which, in general,
renders other observables such as the usual momentum or position as
non-Hermitian operators. The metric depends on one real parameter, the
full range of which is investigated. The explicit functional dependence of the
metric and each associated Hamiltonian is given. A specific choice of this
parameter determines a specific combination of position and momentum
as being an observable; this can be in particular either standard position or
momentum, but not both simultaneously. Singularities of the
metric are explored and their removability is investigated. The physical significance of
these findings is discussed
\end{abstract}

\maketitle

There is continuing interest in the study of non-Hermitian Hamilton operators.
Apart from the obvious situations relating to open systems,
interest is focused upon a specific class of non-Hermitian operators
giving rise to a real spectrum \cite{sgh,bender,mostaf}. 
(See Ref.\cite{JPASpecialIssue} for a recent survey and additional references.)

In Ref. \cite{sgh} reality of the spectrum within the context of a consistent quantum 
mechanical framework is quite generally linked to 
the existence of a positive definite metric operator, giving rise to what is termed 
quasi-hermiticity, while in Ref. \cite{bender} it is conjectured from numerical evidence
that for the class of 
non-Hermitian Hamiltonians studied there, reality of the spectrum results from  
symmetry under simultaneous parity and time
reversal operations (denoted by $\cal{P}$ and $\cal{T}$) -- so-called 
$\cal{P}\cal{T}$-symmetry.
It has subsequently been strictly
proven for particular $\cal{P}\cal{T}$-symmetry cases that the full spectrum 
is in fact real \cite{dorey,shin}. (In Refs. \cite{mostaf} 
the relationship between quasi-Hermiticity \cite{sgh} and $\cal{P}\cal{T}$-symmetry \cite{bender} is explored and 
elucidated in the context of pseudo-Hermiticity.)
 
One particular paradigm
falling into this class is the simple non-Hermitian harmonic oscillator
given by the Hamiltonian
\beq
H=\om  (a^{\dagger }a +\frac{1}{2}) +  \alp a^2 + \beta a^{\dagger 2}  \label{ham}
\eeq
which is manifestly non-Hermitian for $\alp \ne \beta $, but $\cal{PT}$-symmetric 
($\omega$, $\alpha$ and $\beta$ are real parameters). Here we use the
usual boson operators 
\beq
a=\sqrt{\frac{\om  }{2}} \hat x + \frac{i}{ \sqrt{2\om  }}\hat p \label{bos}
\eeq
and correspondingly for $a^{\dagger}$, with $\hat x$ and $\hat p$ being
the usual (Hermitian) position and momentum operators, respectively.

An extensive study of the properties of (\ref{ham}) has been
undertaken in Ref. \cite{swanson}. While the quoted paper is rather implicit,
it has been followed by more explicit investigations \cite{gey,jones,sgplb,sg}.
The emphasis in \cite{gey,sgplb,sg} lies on the non-uniqueness of the metric
with respect to which the non-Hermitian Hamiltonian appears as a
quasi-Hermitian operator, i.e.~Hermitian with respect to a non-trivial metric 
and its associated inner product, {\it viz.}
\beq  
\Theta H=H^{\dagger }\Theta \label{nonh} 
\eeq
with $\Theta $ being a positive Hermitian operator defining the new scalar
product by
$$\langle \cdot |\cdot \rangle _{\Theta }:=\langle \cdot|\Theta \cdot \rangle $$
where $\langle \cdot|\cdot \rangle $ is the usual scalar
product, employing the $L^2$-metric being the identity. Obviously (\ref{nonh}) 
guarantees that the non-Hermitian $H$
is Hermitian with respect to $\langle \cdot|\cdot \rangle _{\Theta }$. 
Moreover, using the positive square root of the metric operator
$\Theta $, the non-Hermitian $H$ can be transformed into a Hermitian operator
 with respect to the $L^2$-metric by the similarity transformation
\beq
h_S=SHS^{-1} \label{hherm}
\eeq
with $S^2=\Theta $, $S$ being likewise positive Hermitian. The essential
point addressed explicitly in the present note
is the non-uniqueness of $S$ and $\Theta $. In fact, various forms have been given in \cite{gey,jones,sgplb,sg}.

In the spirit of a rather general investigation \cite{sgh} about
non-Hermitian Hamiltonians and their associated metric operators we present in this
note a complete analysis of the whole range of operators $S$
yielding Hermitian operators $h_S$ according to (\ref{hherm}) using
(\ref{ham}) for $H$. For the problem at hand, our major finding is a continuous
set of operators $S$ depending on one real parameter. Our emphasis
lies on the physical significance of the specific choice of the metric
in that a particular value of the parameter yields, apart from $h_S$,
a further Hermitian operator (with respect to the $L^2$-metric) being
another observable. Such further observable can be either the position or
momentum operator, but in general a combination thereof, such as for instance
the occupation operator.

We recall that
the spectrum of $H$ is given by $E_n=(n+1/2)\Omega $ with 
$\Omega =\sqrt{\om  ^2-4\alp  \beta}$; of course it must coincide with that of $h_S$.

Guided by specific forms given in \cite{jones,gey,sgplb,sg} we make a
general ansatz for $S$, {\it viz.}
\beq
S=\exp A, \quad A=\ep  a^{\dagger} a+\eta a^2 +\eta^*a^{\dagger 2}
\label{trans}
\eeq
being a positive Hermitian operator as long as $\ep  ^2-4
\eta\eta^*>0$ (the
asterisk denoting complex conjugation); for this to hold $\ep  $ must be real.

Using the expressions
\bea
SaS^{-1}&=&(\cosh \theta -\frac{\ep }{\theta } \sinh \theta ) \;
a-2\frac{\eta^*}{\theta } \sinh \theta \; a^{\dagger} \\
Sa^{\dagger}S^{-1}&=&(\cosh \theta +\frac{\ep }{\theta } \sinh \theta )\;
a^{\dagger }+2\frac{\eta}{\theta } \sinh \theta \; a  \label{trbos}
\eea
with $\theta =\sqrt{\ep  ^2-4 |\eta|^2}$
we obtain
\beq
h_S=SHS^{-1}=U(\ep ,\eta) (a^{\dagger }a +\frac{1}{ 2}) + V(\ep ,\eta)
    a^2+W(\ep ,\eta)a^{\dagger 2} \label{herm}
\eeq
for some $U,V$ and $W$ being obtainable after some algebra; the three
functions depend in fact also on
$\om , \alp ,\beta $. Below explicit expressions are given for $h_S$ for the
whole available range of the parameter $\eta $.

We require $h_S$ to be Hermitian, i.e.~$U$ must be real and
$V=W^*$. This leads to
\beq
\frac{\tanh 2\theta }{\theta }=\frac{\alp -\beta }{(\alp +\beta )\ep  -2
  \om  \eta }  \label{master}
\eeq
and $\eta =\eta ^*$. The transformation (\ref{herm}) invokes a
corresponding inverse transformation for the position and momentum operators
occurring in $H$ \cite{kretsch,mostafbatal}. After suitable rescaling they read
\bea
x=S^{-1}\hat xS&=&
\cosh \theta \; \hat x+\frac{i}{\om  }\frac{\ep  -2 \eta}{\theta } \sinh \theta \; \hat p
\label{coo1} \\
p=S^{-1}\hat pS&=&
\cosh \theta \; \hat p-i \om \, \frac{\ep  +2 \eta}{\theta } \sinh \theta \; \hat x. \label{coo2}
\eea
While $x$ and $p$ are by construction quasi-Hermitian with respect to the metric $\Theta$, 
and hence observables, these expressions clearly show that it is not clear {\it a priori}
whether $\hat x$ or $\hat p$, or a suitable combination  of those, remain observables when
viewed in conjunction with the original Hamiltonian (\ref{ham}). In fact, such property depends on the particular
choice of the metric. In the following we use instead of $\eta $ the parameter $z=\ep  /(2 \eta )$
with $z\in [-1,1]$.

The relation (\ref{master}) covers the whole range of possible parameter values
that determine the metric. For a given set of parameters 
prescribing $H$ (that is $\om ,\, \alp$
and $\beta $) we obtain from (\ref{master}) a relationship between
$z$ and $\ep  $. In other words, the only free parameter that determines the metric is $z$
while $\ep $ is determined by
\beq
\ep =\frac{1}{2\sqrt{1-z^2}}{\rm arctanh}  \frac {(\alp -\beta )\sqrt{1-z^2}}{\alp +\beta -z\om }.
\label{eps}
\eeq
Using the substitutions (\ref{bos},\ref{coo1},\ref{coo2}) and (\ref{eps}) 
slightly tedious but straightforward algebra leads to the Hermitian set of Hamiltonians
\beq
h_{S(z)}=\frac{1}{2}(\mu (z)\,\hat p^2+\nu (z)\,\hat x^2) \label{hs}
\eeq
with
\bea
\mu (z)&=&\frac
{-z(\alp+\beta)+\om -
(\alp+\beta-z\om)
\sqrt
{1-
\frac{(1-z^2)(\alp -\beta )^2} {(\alp +\beta-z\om ) ^2} }} 
{(1+z)\om }  \label{mu}  \cr
\nu (z)&=&-\om \frac
{z(\alp+\beta)-\om -
(\alp+\beta-z\om)
\sqrt
{1-
\frac{(1-z^2)(\alp -\beta )^2} {(\alp +\beta-z\om ) ^2} }} 
{1-z }  .  \label{vu}
\eea
The similarity transformation (\ref{trans})
that gives rise to (\ref{hs}) from (\ref{herm}) is obtained
in a similar vein
\bea
S(z)&=&\bigg(\frac{\alp+\beta-\om z+(\alp -
 \beta)\sqrt{1-z^2}}{\alp+\beta-\om z-(\alp -\beta)\sqrt{1-z^2}}\bigg)
 ^{\frac{1}{4\sqrt{1-z^2}}
(a{^\dagger }a+\frac{z}{2}(a^2+a^{\dagger 2}))}               \cr
 &=&\bigg(\frac{\alp +\beta-\om z+(\alp -\beta)\sqrt{1-z^2}}{\alp
    +\beta-\om z-(\alp -\beta)\sqrt{1-z^2}
}\bigg)^{\frac{1}{8\om \sqrt{1-z^2}} (p^2(1-z)+\om ^2 x^2(1+z)-\om )}.
\label{simz}
\eea

Specific cases have been given in  \cite{jones,gey,sgplb,sg}:
\begin{itemize}
\item{(i)} for $z =0$
yielding from (\ref{eps}) $\ep  =1/4\log (\alp /\beta )$ and thus 
\beq
\Theta =S^2=\bigg(\frac{\alp }{\beta }\bigg) ^{\frac{\hat N}{2}} \label{t1}
\eeq
and
\beq
h_{S(z =0)}=\frac{\om  -2\sqrt{\alp \beta}}{2 \om  }\; \hat
p^2+\frac{\om  }{2} (\om  +2\sqrt{\alp \beta}) \; \hat x^2, \label{h1}
\eeq

\item{(ii)} for $z =1$ yielding $\ep  =-(\alp
-\beta)/(2(\om -\alp -\beta))$ and thus
\beq
\Theta =S^2=\exp \bigg(-\frac{\alp-\beta  }{\om -\alp -\beta } \; \om \hat x ^2
\bigg) \label{t2}
\eeq
and
\beq
h_{S(z=1)}=\frac{\om -\alp-\beta}{2 \om  }\;
  \hat p^2+\frac{\om  \Omega ^2}{2(\om -\alp
    -\beta ) }\; \hat x^2 \label{h2}
\eeq

\item{(iii)} for $z=-1$ yielding $\ep  =(\alp
-\beta )/(2(\om +\alp +\beta))$ and thus
\beq
\Theta =S^2=\exp \bigg(\frac{\alp-\beta  }{\om +\alp +\beta }\; \frac{\hat p ^2}{\om } \bigg) \label{t3}
\eeq
and
\beq
h_{S(z=-1)}= \frac{\Omega ^2}{2 \om  (\om 
  +\alp+\beta)}\; \hat p^2+
\frac{\om  (\om 
+\alp+\beta)}{2}\; \hat x^2 . \label{h3} 
\eeq

\end{itemize}

We have presented $h_S$, that is the hermitized forms of $H$, 
in (\ref{hs}) and their special forms in (\ref{h1}), (\ref{h2}) and (\ref{h3}) 
in terms of the traditional momentum and position operators to indicate that they
have all the same spectrum; they are simply rescaled forms of each
other. In fact, while this is obvious by inspection from (\ref{h1}), (\ref{h2}) and (\ref{h3}), 
the general form (\ref{hs}) obeys as well identically the relation 
$\mu \nu =\Omega ^2=\om ^2-4 \alp \beta $, as it should.
However, according to (\ref{coo1}) and (\ref{coo2})
the metric associated with a particular choice of $z$ does not -- using the $L^2$-metric -- yield 
Hermitian position and momentum
operators. It does though yield the Hermitian combination
\beq
O=\om ^2 x^2 (1+z)+p^2 (1-z) \label{oo}
\eeq
which is -- as we conclude from (\ref{coo1}) and (\ref{coo2}) -- 
identical to the manifestly $L^2$-Hermitian operator
$$\hat O=\om ^2 \hat x^2 (1+z)+\hat p^2 (1-z) . $$
Note that $O=\hat O$ is Hermitian with respect to both the $L^2$-metric, 
being the identity, and the most general metric $\Theta (z)$ (compare also the final example 
in Ref. \cite{kretsch}).
Note further that $z=0$ implies $O\sim \hat N=a^{\dagger }a$, the number operator.
In contrast, $z=1$ yields, according to (\ref{coo1})
and (\ref{coo2}), a metric for which $x$ is $L^2$-Hermitian but $p$ is not. In fact, $S$
and thus $\Theta $ is now a function of $\hat x$ only and we read from
(\ref{coo1})
$$x=S^{-1}\hat x S= \hat x. $$
{\it Mutatis mutandis} $z =-1$ gives a non-Hermitian $x$
but the Hermitian momentum
$$p=S^{-1}\hat p S= \hat p. $$

These results nicely demonstrate the point made in \cite{sgh}, and recently 
elaborated in \cite{zg,mostafozcelik}, in that
the metric can be made unique by choosing, or constructing, further 
operators as
observables ({\em i.e.~operators being quasi-Hermitian with respect to the same metric}) 
to form an {\em irreducible set} comprising the
Hamiltonian. The examples discussed in detail specify one more operator to be
chosen, that is (i) the number or (ii) the position or (iii) the momentum operator.

While  the specific choices  made  for  $z$ may  be  physically
appealing  as one  of each  choice allows  at least  one of  the three
operators ($\hat  N, \hat x, \hat p$) to  be an observable in conjunction with 
the non-Hermitian Hamiltonian (\ref{ham}), any  other choice of
$z \in[-1,1]$ may  be possible in principle. Such
other choice yields, however,  another Hermitian  combination of
the  momentum and  position operator as given in (\ref{oo}).  Whether such combination has any
particular  physical meaning  had to  be judged  by the  specific case
considered.

In turn, depending on the choice of parameters for $H$ (while duly
observing $\om  ^2\ge 4 \alp \beta$), there may be
combinations that don't allow a real solution for
$\ep  $ of (\ref{eps}) even if $z$ is properly chosen in the interval
$[-1,1]$. In fact, the obvious requirement that the argument of the hyperbolic
arctanh is not greater than unity -- which is equivalent to the square root occurring in
(\ref{mu}) being real -- reveals that there is no real solution for
$z\in [z_-,z_+]$ with
\beq
z_{\pm}=\frac{(\alp +\beta)\om  \pm(\alp -\beta)\Omega}{\om 
^2+(\alp -\beta )^2}.  \label{zcrit}
\eeq
The numerical example $\om  =1, \alp=1/2,
\beta=1/4$ yields $[0.54\ldots,0.87\ldots]$ as the disallowed region for $z$.
Note that $z_+=1$ for $\om  =\alp +\beta $. This combination is
obviously incompatible with the choice $z =1$ as seen
from (\ref{t2}) and (\ref{h2}).  In other words, for 
$\om  =\alp +\beta $ ($\alp \ne \beta $) the position operator simply cannot be Hermitian.
We stress that as $h_S$ fails to be Hermitian when $z\in [z_-,z_+]$, the metric
$S$ is ill defined for these values of $z$ as the argument to be exponentiated
in (\ref{simz}) is negative. The metric is singular (infinity) at $z=z_-$ and zero at $z=z_+$. 

The singularity just described of the metric is spurious, however. It means that it is removable \cite{sg} by
making another choice for the metric, yet at the expense of trading in singularities elsewhere. For the
present problem this is achieved by simply making the replacement $z\to -z$ everywhere. This 
entails in particular that
\begin{itemize}
\item{} in (\ref{mu}) $\mu(z)$ is to be replaced by $\mu(-z)$,
  $\nu(z)$ by $\nu(-z)$ and in (\ref{simz}) $S(z)$ by $S(-z)$
\item{} the region where the metric is ill defined is now at $z\in [-z_+,-z_-]$
\item{} $x\equiv \hat x$ for $z=-1$ with $p$ non-Hermitian
\item{} $p\equiv \hat p$ for $z=+1$ with $x$ non-Hermitian
\item{} item(ii) leading to (\ref{t2}) and (\ref{h2}) must now read
\begin{itemize}
\item{(ii)} for $z =-1$ \\ -- with the expressions following remaining unchanged
\end{itemize}
\item{} item(iii) leading to (\ref{t3}) and (\ref{h3}) must now read
\begin{itemize}
\item{(iii)} for $z =1$ \\ -- with the expressions following remaining unchanged
\end{itemize}
\item{} (\ref{oo}) now reads $O=\om ^2 x^2 (1-z)+p^2 (1+z)$ and correspondingly for $\hat O$.
\end{itemize}

It is worth mentioning that the singularities of the metric persist
if the parameters of the Hamiltonian are chosen such that $z_+$ and
$z_-$ coincide. Using (\ref{zcrit}) this happens when $\Omega =0$ -- ignoring
the trivial case $\alp =\beta $ --, that is at an exceptional point \cite{kato,hs,heiss},
where all energies coalesce. 
With $\om =2\sqrt{\alp \beta}$ the expression reads for $S(z)$
\beq
S(z) =\bigg(\frac{\alp +\beta-2\sqrt{\alp \beta} z+(\alp -\beta)\sqrt{1-z^2}}{\alp
    +\beta-2\sqrt{\alp \beta } z-(\alp -\beta)\sqrt{1-z^2}
}\bigg)^{\frac{1}{8\om \sqrt{1-z^2}} (p^2(1-z)+\om ^2 x^2(1+z)-\om
  )}.  \label{simsp}
\eeq
When $z\to z_+$ the denominator of (\ref{simsp}) vanishes to second
order. The metric is singular at the exceptional point, which in a more general
situation would be indicative of a
phase transition \cite{heiss,sg,heissjpa}.

Having completely analysed the Hamiltonian considered there remains
the question: what choice to make to obtain unique physical answers?
In this context we stress that, while the hermitized Hamiltonians have
the same spectrum, the corresponding wave functions do depend on 
$z$. In fact, the set of Hamiltonians (\ref{hs}) 
clearly yield the well known harmonic oscillator wave
functions but with distinctly different arguments for the Gaussian and
Hermite polynomials, the respective arguments being given by the combination
$(\nu /\mu )^{1/4} x$.
In other words, not only is the preferred choice of the additional observable a
function of $z$ and thus of the metric, but so are other variables
such as transition matrix elements \cite{mus}. We are not aware of a
convincing argument that could fix the choice of the metric. There
remains an ambiguity. Any further elucidation should come from a
genuine physical situation that is described by a $\cal{PT}$-symmetric
or other non-Hermitian Hamiltonian which is quasi-Hermitian, with a
real spectrum. If it exists, Nature will tell what metric
she prefers under given circumstances.

{\it Note added in proof:} Using arguments based on a perturbative
approach it has been
suggested \cite{ali} that the classical limit of the hermitized
Hamiltonian be independent of the particular choice of the metric. Our
nonperturbative results do not support this suggestion. In fact, the
quantities $\mu (z)$ and $\nu (z)$ in (\ref{vu}) explicitly depend on
$z$ and so does the Hamiltonian in (\ref{hs}). While the oscillator
frequency is of course independent of $z$, the mass term becomes $1/\mu (z)$
and the classical energy $E_{\rm cl}=A^2 \Omega ^2/(2\mu(z))=\nu (z)A^2/2$
($A$=amplitude of the classical oscillation). The (spurious)
singularities at $z_{\pm }$ given in (\ref{zcrit}) also appear in the
mass term; the mass and classical energy remain finite at
$z_{\pm }$ but they are complex for $z\in [z_-,z_+]$. We note that the metric
operator has an essential singularity in the classical limit ($\hbar
\to 0$), that is it cannot be expanded in powers of $\hbar $.

{\bf Acknowledgement}
 
We thank Frederik Scholtz for a critical reading of the manuscript and for numerous illuminating discussions.


\begin{thebibliography}{99}

\bibitem{sgh} Scholtz F G, Geyer H B and Hahne F J W 1992, Ann. Phys. (N.Y.) {\bf 213} 74
\bibitem{bender} Bender C M and Boettcher S 1998, Phys. Rev. Lett. {\bf 80} 4243
\bibitem{mostaf} Mostafazadeh A 2002, J. Math. Phys. {\bf 43} 205; Mostafazadeh A 2002, J. Math. Phys. {\bf 43} 2814; 
               Mostafazadeh A 2002, J. Math. Phys. {\bf 43} 3944
\bibitem{JPASpecialIssue} Geyer H B, Heiss W D and Znojil M (Eds.) 2006, J. Phys. A: Math. Gen {\bf 39}, No. 32 
         (Special issue on {\it The Physics of Non-Hermitian Operators})
\bibitem{dorey} Dorey P, Dunning C and Tateo R 2001, J. Phys. A: Math Gen. {\bf 34} 5679
\bibitem{shin} Shin K C 2002, Commun. Math. Phys. {\bf 229} 543 
\bibitem{swanson} Swanson M S 2004, J. Math. Phys. {\bf 45} 585
\bibitem{gey} Geyer H B, Snyman I and Scholtz F G 2004, Czech. J. Phys. {\bf 54} 1069
\bibitem{jones} Jones H F 2005, J. Phys. A: Math. Gen. {\bf 38} 1741
\bibitem{sgplb} Scholtz F G, Geyer H B 2006, Phys. Lett. B  {\bf 634} 84
\bibitem{sg} Scholtz F G, Geyer H B 2006, J. Phys. A: Math. Gen. {\bf 39} 10189
\bibitem{kretsch} Kretschmer R and Szymanowski L 2004, Czech. J. Phys. {\bf 54} 71
\bibitem{mostafbatal} Mostafazadeh A and Batal J 2004, J. Phys. A: Math. Gen. {\bf 37} 11645
\bibitem{zg} Znojil M, Geyer H B 2006, Phys. Lett. B  {\bf 640} 52
\bibitem{mostafozcelik} Mostafazadeh A and Ozcelik S 2006, quant-ph/0607120
\bibitem{kato} Kato T 1976, {\it Perturbation Theory for Linear Operators} 2nd edn (Heidelberg: Springer)
\bibitem{hs} Heiss W D and Sannino A L 1990, J. Phys. A: Math. Gen. {\bf 23} 1167
\bibitem{heiss} Heiss W D 2004, Czech. J. Phys. {\bf 54} 1091
\bibitem{heissjpa} Heiss W D 2006, J. Phys. A: Math. Gen. {\bf 39} 10081
\bibitem{mus} Musumbu D P 2006, MSc thesis, University of Stellenbosch
  (unpublished)
\bibitem{ali} Mostafazadeh A 2006,  J. Phys. A: Math. Gen. {\bf 39} 10171

\end{thebibliography}
\end{document}